# Accelerating Loading WebGraphs in ParaGrapher

Mohsen Koohi Esfahani
0000-0002-7465-8003

*Abstract*—ParaGrapher is a graph loading API and library that enables graph processing frameworks to load large-scale compressed graphs with minimal overhead. This capability accelerates the design and implementation of new high-performance graph algorithms and their evaluation on a wide range of graphs and across different frameworks.

However, our previous study identified two major limitations in ParaGrapher: inefficient utilization of high-bandwidth storage and reduced decompression bandwidth due to increased compression ratios. To address these limitations, we present two optimizations for ParaGrapher in this paper.

To improve storage utilization, particularly for high-bandwidth storage, we introduce ParaGrapher-FUSE (PG-Fuse) a filesystem based on the FUSE (Filesystem in User Space). PG-Fuse optimizes storage access by increasing the size of requested blocks, reducing the number of calls to the underlying filesystem, and caching the received blocks in memory for future calls.

To improve the decompression bandwidth, we introduce CompBin, a compact binary representation of the CSR format. CompBin facilitates direct accesses to neighbors while preventing storage usage for unused bytes.

Our evaluation on 12 real-world and synthetic graphs with up to 128 billion edges shows that PG-Fuse and CompBin achieve up to 7.6 and 21.8 times speedup, respectively.

## I. INTRODUCTION

Graph loading libraries [1–3] and synthetic graph generators [4–7] play a crucial role in designing high-performance graph algorithms, which often rely on experimental evaluation. Specifically, they have three key impacts: (i) accelerating the implementation of graph algorithms, (ii) enabling the evaluation of new algorithms across a diverse range of graph datasets, and (iii) facilitating fast and straightforward evaluation of graph algorithms across different frameworks.

We previously introduced ParaGrapher [1], a graph loading API and library designed to efficiently load the large-scale graphs compressed in WebGraph format [8]. ParaGrapher offers flexible loading options, allowing the users to load the entire graph or some partitions of the graph, either synchronously (blocking) or asynchronously (non-blocking). This flexibility enables integration with shared-memory, out-of-core, and distributed-memory graph frameworks.

Our evaluation demonstrated ParaGrapher's strength in loading compressed graphs in comparison to other graph formats, particularly for loading from low bandwidth storage, such Hard Disk Drives (HDD). However, when utilizing high bandwidth storage, such as Solid Stated Drives (SSDs), we observed that decompression becomes a bottleneck, limiting the achievable bandwidth and preventing the system from fully benefiting from the device's high bandwidth capabilities.

In this paper, we address the limited decompression bandwidth in ParaGrapher by presenting two optimizations. Our first optimization stems from the observation that the Java-based WebGraph implementation exhibits a pattern of frequent, small blocks (128 kB) storage accesses. This behavior leads to three key issues (i) reduced read bandwidth, (ii) increased latency, and (iii) inefficient utilization of the internal prefetcher in distributed file systems, such as Lustre, which are commonly found in supercomputing environments.

To mitigate this issue, we designed and implemented ParaGrapher-FUSE (PG-Fuse), a caching filesystem that retrieves large-sized blocks of data (e.g., 32 MiB) from the underlying storage and caches them in memory for future use. We leveraged the Filesystem in User Space (FUSE) to implement the PG-Fuse.

Our second optimization targets reducing the decompression computational overhead through light-weight compression. We introduce CompBin a compact binary representation of Compressed Sparse Row/Column (CSR/CSC) [9]. CompBin removes the unused bytes in storing the vertex ID. Specifically, for a graph with $|V|$ vertices, CompBin uses $\lceil (log_2|V|)/8 \rceil$ bytes to represent each vertex ID, enabling efficient storage and random access to the neighbors. Furthermore, decompression is efficiently performed using only a few shift and add operations, resulting in reduced computational overhead.

Our evaluation on 12 real-world and synthetic graphs, ranging in size up to 128 billion edges, demonstrates that PG-Fuse yields a speedup of 0.9–7.6 times. CompBin achieves a speedup of up to 21.8 times, mainly efficient for graphs characterized by a small size.

In summary, the paper makes these key contributions:
- Introducing PG-Fuse, a caching filesystem designed to enhance the read bandwidth of ParaGrapher,
- Introducing CompBin, a compact binary representation of the CSR format for storing graphs, and
- Evaluating PG-Fuse and CompBin.

The paper is continued with a review of background materials in Section II. We introduce PG-Fuse in Section III and CompBin in Section IV. The evaluation is presented in Section V and Section VI concludes the discussion.

## II. BACKGROUND & RELATED WORK

A graph $G = (V, E)$ has a set of vertices $V$, and a set of directed edges $E$. The Compressed Sparse Row (CSR) or Column (CSC) [9] consists of two arrays: an *offsets* array containing $|V| + 1$ elements, and a *neighbors* array of $|E|$ elements. The offsets array is indexed by a vertex ID and specifies the index of the first neighbor of that vertex in the neighbors array. The neighbors array contains the vertex ID of the source or destination endpoint of the edges.

*A. ParaGrapher*

Graph compression is vital for the efficient storage and transferring of graph datasets. WebGraph[1] [8] is a graph compression and processing framework. WebGraph has been used for compressing web [10, 11], version-control [12, 13] and sequence similarity [14] graphs.

ParaGrapher [1] is an API and library designed for efficient loading of large-scale compressed WebGraphs in shared-memory, distributed-memory, and out-of-core graph processing frameworks developed in C and C++. The WebGraph framework has been implemented in Java and, more recently, in Rust [15]. ParaGrapher utilizes the Java implementation of WebGraph and employs a consumer-producer pattern for loading graphs compressed in WebGraph format. In this architecture, the consumer side is implemented in the C language, while the producer side is implemented in Java.

To enable communication and synchronization between the two sides, ParaGrapher allocates shared memory that can be accessed by both processes. The C-side program is responsible for allocating memory for the shared reusable buffers. The Java-side program writes to these buffers and once the data is passed to the C side, it is then forwarded to the user through the user-defined callback functions, allowing the user to manage the preferred memory system in the graph framework.

The source code of ParaGrapher is available online[2]. By leveraging ParaGrapher, researchers and developers can potentially save at least 1,500 Lines of code. The API documentation is presented online on the ParaGrapher's repository[3].

*B. Related Work*

Other compression methods include identifying bicliques to reduce the number of edges that should be stored [16–18], graph summarization to create a group of vertices containing edges [19, 20], rule-based compression [21], and lossy compression methods such as frontier sampling [22], query-preserving compression [23], and importance-based sampling [24]. Compression surveys are available on [25–27].

PIGO [2] is a library for parallel loading uncompressed graphs. GVEL [3] optimizes conversion of edge lists in coordinated format to CSX. EndGraph [28] optimizes load-balance and sorting in the preprocessing step of distributed graph computing. GraPU optimizes streaming preprocessing [29]. Then et al. present an evauation of graph loading methods [30]. LV et al. present a survey of graph preprocessing methods [31].

## III. PARAGRAPHER-FUSE

Our previous evaluation of ParaGrapher [1] revealed that decompression overhead limits ParaGrapher's output bandwidth, highlighting the need for optimization techniques to improve the decompression process.

[1]https://webgraph.di.unimi.it/
[2]https://github.com/dipsa-qub/ParaGrapher
[3]https://github.com/dipsa-qub/ParaGrapher/wiki/API-Documentation

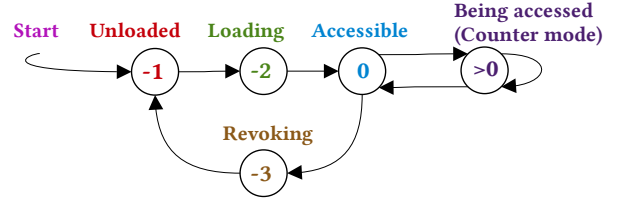

Fig. 1: State transition diagram of block status in PG-Fuse

Our experiments identified that frequent and small granularity of storage accesses are a primary source of decompression inefficiency. We observed that the Java Virtual Machine requests blocks of up to 128 kB, resulting in inefficient utilization of the underlying storage, particularly, for high-bandwidth storage devices such as SSDs and distributed file systems. Furthermore, distributed file systems like Lustre are equipped with prefetchers to read ahead and optimize performance [32, 33]. However, the prefetcher may not function effectively when dealing with small requested blocks.

To address this issue, we considered two potential solutions: (i) optimizing the read process in WebGraph library and (ii) designing a caching filesystem that utilizes large granularity in read accesses. The first solution would introduce a dependency between ParaGrapher and specific versions of WebGraph, potentially limiting the direct utilization of future WebGraph versions in ParaGrapher. In contrast, the second solution is fully independent of the WebGraph implementation, making it a durable solution.

We implemented the filesystem using the Filesystem in User Space (FUSE)[4] library, and named it ParaGrapher-FUSE (PG-Fuse). PG-Fuse divides an inode's total capacity into large blocks, with a default size of 32 MiB. When a read request is received, PG-Fuse reads the requested block and stores it in memory. Subsequent reads from the same block are answered from the cache, eliminating the need to access the underlying filesystem. To manage memory efficiently, PG-Fuse tracks the last access time for each block. This allows PG-Fuse to revoke recently-unused blocks from its cache.

To accommodate concurrent thread accesses, PG-Fuse assigns an integer status value to each block and protects status variables by atomic memory accesses. A block may be in the following states:

- 0: The block is loaded and accessible by threads.
- Positive integer values: The number of concurrent threads accessing the block. In this state, the status variable serves as a counter.
- -1: The block has not been loaded.
- -2: A thread is currently loading the block's contents. Other threads should wait before accessing it.
- -3: The block is being revoked by a thread.

Figure 1 illustrates the transition of block status in PG-Fuse.

When opening a graph for reading, users can pass an argument to ParaGrapher to request the use of PG-Fuse. Para-

[4]https://github.com/libfuse/libfuse

Grapher then mounts the graph files in the PG-Fuse filesystem. When the user requests to close the graph, ParaGrapher unmounts the PG-Fuse filesystem and releases all allocated memory for its internal data structures and non-expired blocks.

Section V-B evaluates PG-Fuse when mounting on top of a Lustre filesystem and shows that PG-Fuse could achieve a speedup of up to 7.6 times.

## IV. CompBin

The other strategy to reduce decompression overhead is to utilize light-weight graph compression algorithms with lower/greater compression ratios [26, 34]. We introduce CompBin a compact binary representation of the CSR format. The binary representation of CSR has been widely adopted in graph frameworks. This format allocates a constant number of bytes (often 4 bytes) for each vertex ID in the $neighbors$ array and offers the advantage of directly mapping the $neighbors$ array to memory using `mmap()` system call.

In CompBin, we address the inefficient utilization of the storage space and bandwidth in binary CSR format for assigning the same number of bytes for all datasets. To mitigate these issues, CompBin represents vertex IDs in the $neighbors$ array using the minimum number of bytes required. Specifically, for a graph with $|V|$ vertices, CompBin allocates $b = \lceil (log_2|V|)/8 \rceil$ bytes for storing a vertex ID.

Compared to binary CSR, CompBin reduces storage size while maintaining direct access to the $neighbors$ array. Assuming a $neighbors$ array with 1 byte elements (i.e., of type `uint8_t`), its size is $b|E|$, and the vertex ID of the $n$-th neighbor of vertex $v$ is calculated using the following formula.

$$\sum_{i=0}^{i=b-1} neighbors[(offsets[v] + n) * b + i] << (8i) \quad (1)$$

As a result, the decompression of CompBin is performed using a few shift and add operations, while preserving the other benefits of the binary CSR format. Notably, for $2^{24} \leq |V| < 2^{32}$, the CompBin representation is equivalent to the binary CSR format.

Section V-C demonstrates that CompBin is mainly efficient for small graphs in our datasets, achieving a speedup of up to 21.8 times in comparison to ParaGrapher for loading graphs.

## V. Evaluation

### A. Environmental Setup

We conducted experiments on a machine with two AMD 7702 CPUs, totaling 128 cores with a frequency of 2–3.4 GHz, 512 MB L3 caches, 2 TB of memory, and running CentOS 8. Hyper-threading is disabled and the machine is connected to a 2 Petabytes Lustre filesystem with a SSD pool. The filesystem is shared among users of the cluster.

The characteristics of the graph datasets used in our experiments are summarized in Table I. Our datasets comprise a diverse range of graph types, including web graphs [8, 10, 11, 35–38], social networks [39], synthetic graphs [40], version control history graphs (VCH) [12, 13], and bio graphs [14].

TABLE I: Datasets

| Name | |V| | |E| | Type | Size on Storage (GiB) | |
|---|---|---|---|---|---|
| | | | | WebGraph | CompBin |
| enwiki-2023 | 6.6M | 165.2M | Web | 0.3 | 0.5 |
| twitter-2010 | 41.7M | 1.5G | Social | 2.5 | 5.8 |
| sk-2005 | 50.6M | 1.9G | Web | 0.5 | 7.6 |
| MS1 | 43.1M | 2.7G | Bio | 5.9 | 10.2 |
| clueweb09 | 1.7G | 7.9G | Web | 12.2 | 42.1 |
| g500 | 536.9M | 17.0G | Synth. | 49.7 | 67.3 |
| gitlab-all | 1.1G | 27.9G | VCH | 14.3 | 112.1 |
| gsh-2015 | 988.5M | 33.9G | Web | 9.2 | 133.6 |
| uk-2014 | 787.8M | 47.6G | Web | 8.2 | 183.2 |
| eu-2015 | 1.1G | 91.8G | Web | 14.1 | 349.9 |
| MSA50 | 1.8G | 125.3G | Bio | 385.2 | 479.9 |
| wdc12 | 3.6G | 128.7G | Web | 57.4 | 506.1 |

The last two columns of Table I show the storage size of graphs in WebGraph format and in CompBin format in Gigabytes. Notably, CompBin uses 3 bytes per vertex ID for `enwiki-2023`, whereas the remaining datasets require 4 bytes per vertex ID, making CompBin equivalent to binary CSR format (Section IV).

This equivalence allows the CompBin results for graphs larger than `enwiki-2023` to be considered representative of binary CSR format in the following sections, enabling a comparison between WebGraph and binary CSR formats.

The source code is compiled with `gcc` 14.0.1 with `-O3` flag, `OpenJDK` 17.0.10, and `WebGraph` 3.6.10.

### B. PG-Fuse

Figure 2 compares the graph loading time of ParaGrapher with and without using PG-Fuse. The results show that PG-Fuse achieves a speedup of 1.3–2.4 for the web graphs, with the maximum value obtained for the smallest graph, `enwiki-2023`, and the lowest speedup for `eu-2015`, the graph with the greatest compression ratio.

For `twitter-2010`, using PG-Fuse results in an approximately 10% performance loss. For small graphs, a large block size of PG-Fuse restricts parallelism for loading the compressed graph from the underlying filesystem, causing all threads to experience an initial delay due to concurrently loading the entire graph. When this delay is not offset by the speedup provided by PG-Fuse for future accesses, the graph loading may lead to performance loss. Reducing the size of the PG-Fuse block for small graphs increases the load parallelism and accelerates the overall loading process.

For bio graphs, `MS1` and `MSA50`, PG-Fuse facilitates 1.9 and 7.6 times speedup. In particular, `MSA50` is the graph with the largest size on storage (Table I). For `g500` as a synthetic graph, PG-Fuse achieves 3.4 times speedup and the speedup for `gitlab-all` control version history graph is 1.7 times. In total, PG-Fuse achieves 0.9–7.6 times speedup.

### C. CompBin

Figure 3 compares the speedup in graph loading time for CompBin and PG-Fuse over the graph loading time in ParaGrapher without PG-Fuse. For `enwiki-2023`, CompBin assigns 3 bytes per vertex ID, achieving a 21.8 times speedup in comparison to ParaGrapher without PG-Fuse and a 2.4 times speedup over ParaGrapher with PG-Fuse.

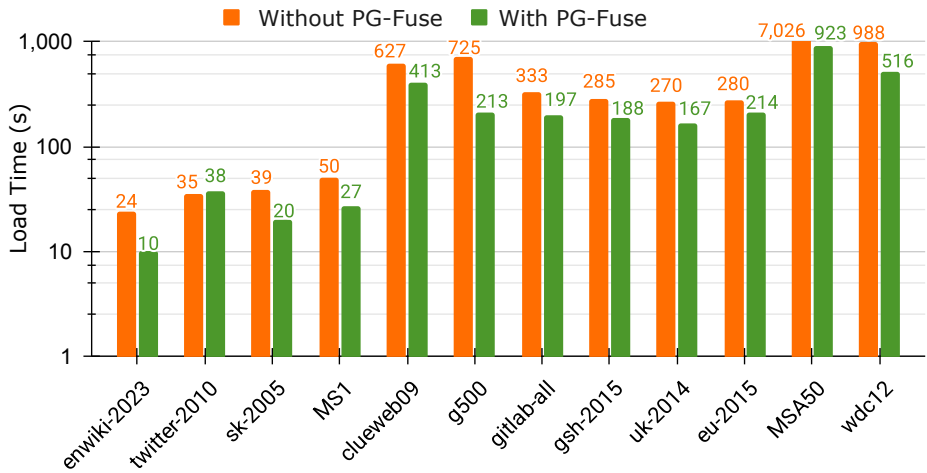

Fig. 2: Graph loading time in ParaGrapher using PG-Fuse and without using PG-Fuse. Values are in seconds.

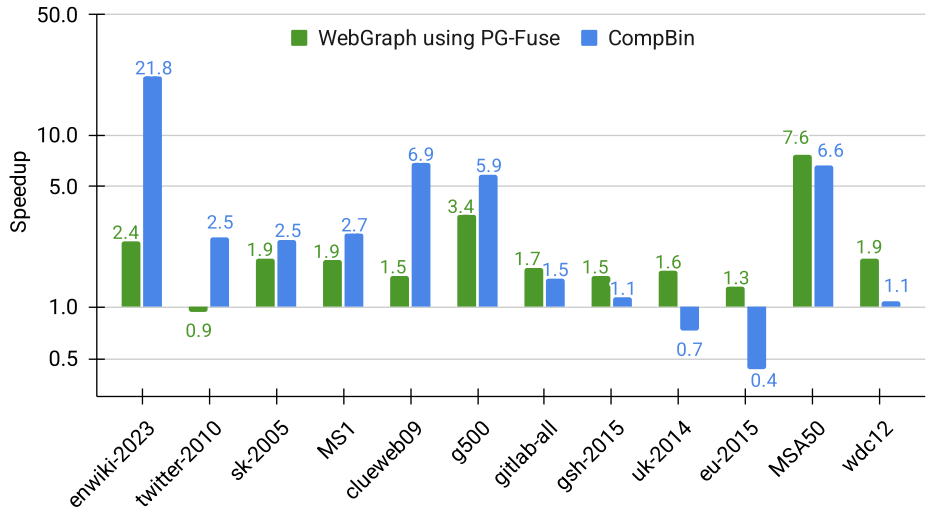

Fig. 3: Speedup of graph loading time for CompBin and PG-Fuse against ParaGrapher without PG-Fuse. CompBin is equivalent to the binary CSR format for graphs larger than enwiki-2023.

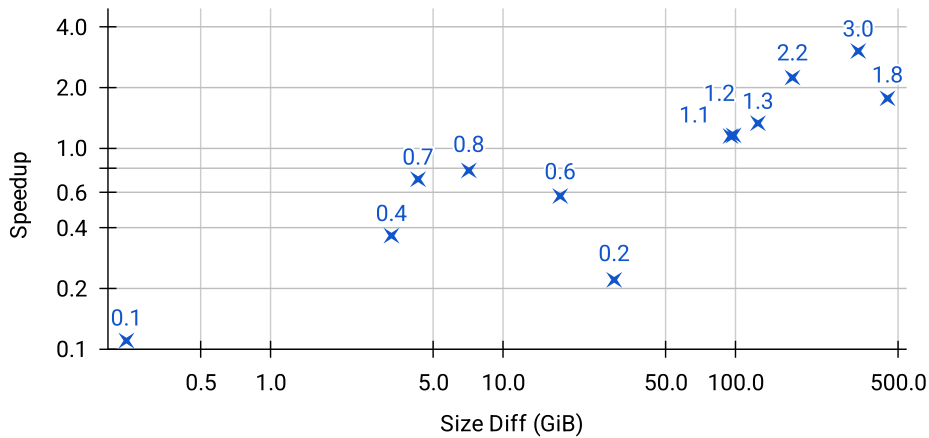

Fig. 4: Comparing PG-Fuse against CompBin for graph datasets. The X-axis shows the difference in graph sizes and the Y-axis shows the speedup of PG-Fuse against CompBin. Values greater than 1 on the Y-axis indicate better performance for PG-Fuse, while values less than 1 indicate better performance for CompBin/binary CSR.

For graphs larger than `enwiki-2023`, CompBin needs 4 bytes per vertex ID and is equivalent to binary CSR format. As shown in Figure 3, for graphs smaller than `g500`, CompBin/binary CSR outperforms ParaGrapher regardless of whether PG-Fuse is used. This is because the small graph size, relative to the storage bandwidth, results in decompression overhead in WebGraph format which is not offset by the reduced read time from the storage.

For graphs larger than `g500`, loading from CompBin/binary CSR is slower than using PG-Fuse for decompressing WebGraph. For large web graphs with high compression ratios such as `uk-2014` and `eu-2015`, CompBin/binary CSR loading becomes storage-bandwidth limited and cannot compete with WebGraph decompression in ParaGrapher, even when PG-Fuse is not used.

### D. PG-Fuse vs. CompBin

To better compare PG-Fuse and CompBin/binary CSR, Figure 4 illustrates the relationship between speedup and storage size difference for various graphs. The Y-axis represents the speedup of the graph loading for PG-Fuse over CompBin and the X-axis shows the difference in storage size for the WebGraph and CompBin formats. Note that CompBin is equivalent to binary CSR format for graphs larger than `enwiki-2023` (point [0.2, 0.1] in the plot).

As shown in Figure 4, when the storage size difference is less than 50 GiB, the speedup is less than 1, indicating faster graph loading in CompBin/binary CSR formats. In contrast, when the storage size difference approaches or exceeds 100 GiB, PG-Fuse outperforms CompBin/binary CSR. It is worth noting that the threshold values of 50 and 100 GiB are dependent on both the storage bandwidth and the computational power of the system and may vary accordingly.

## VI. CONCLUSION & FUTURE WORK

In this paper, we present PG-Fuse, an extension to ParaGrapher that accelerates the loading of compressed graphs in WebGraph format. PG-Fuse optimizes bandwidth utilization of the underlying filesystem, improving the loading and decompression performance.

To minimize decompression overhead, we introduce CompBin, a compact binary representation of CSR format that enables lightweight decompression while allowing direct access to the neighbors array.

Our evaluation demonstrates that PG-Fuse achieves a speedup of up to 7.6 times, particularly beneficial for large graphs, and CompBin achieves a speedup of up to 21.8 times, primarily advantageous for small graphs.

We consider the following cases as future areas of study.

- To further optimize the decompression process, PG-Fuse can be enhanced to track the access pattern of threads and trigger prefetching of subsequent blocks from the underlying filesystem.
- As discussed in Section V-B, adjusting the block size in PG-Fuse can improve its performance for small graphs.
- Considering the speedup achieved by CompBin for small graphs, ParaGrapher can be further optimized by implementing hybrid loading policies to select the fast format based on the graph characteristics.

## CODE AVAILABILITY

The source code for converting WebGraph format to CompBin format is available on https://github.com/MohsenKoohi/WG2CompBin. ParaGrapher is available on https://github.com/MohsenKoohi/ParaGrapher.

## ACKNOWLEDGEMENTS

This work was partially supported by NI-HPC (UKRI EPSRC grant EP/T022175/1).

We used a non-training version of Llama 3.3 to enhance grammar and editing in this paper.

## REFERENCES

[1] M. Koohi Esfahani, M. D'Antonio, S. I. Tauhidi, T. S. Mai, and H. Vandierendonck, "Selective parallel loading of large-scale compressed graphs with ParaGrapher," *CoRR*, 2024. [Online]. Available: https://arxiv.org/abs/2404.19735

[2] K. Gabert and U. V. Çatalyürek, "PIGO: A parallel graph input/output library," in *2021 IEEE International Parallel and Distributed Processing Symposium Workshops (IPDPSW)*. IEEE, 2021, pp. 276–279.

[3] S. Sahu, "GVEL: Fast graph loading in edgelist and compressed sparse row (csr) formats," 2023.

[4] D. Chakrabarti, Y. Zhan, and C. Faloutsos, "R-mat: A recursive model for graph mining." in *SDM*. SIAM, 2004, pp. 442–446.

[5] D. A. Bader and K. Madduri, "Gtgraph: A synthetic graph generator suite," *Atlanta, GA*, vol. 38, 2006.

[6] H. Park and M.-S. Kim, "Trilliong: A trillion-scale synthetic graph generator using a recursive vector model," in *Proceedings of the 2017 ACM International Conference on Management of Data*, ser. SIGMOD '17. ACM, 2017, p. 913–928.

[7] V. Anand, P. Mehrotra, D. Margo, and M. Seltzer, "Smooth kronecker: Solving the combing problem in kronecker graphs," ser. GRADES-NDA'20. ACM, 2020.

[8] P. Boldi and S. Vigna, "The webgraph framework i: Compression techniques," in *Proceedings of the 13th International Conference on World Wide Web*, ser. WWW '04. ACM, 2004, p. 595–602.

[9] Y. Saad, "Sparskit: a basic tool kit for sparse matrix computations - version 2," 1994. [Online]. Available: https://citeseerx.ist.psu.edu/viewdoc/summary?doi=10.1.1.41.3853

[10] P. Boldi, B. Codenotti, M. Santini, and S. Vigna, "Ubicrawler: A scalable fully distributed web crawler," *Softw. Pract. Exper.*, vol. 34, no. 8, p. 711–726, Jul. 2004.

[11] P. Boldi, A. Marino, M. Santini, and S. Vigna, "Bubing: Massive crawling for the masses," *ACM Trans. Web*, vol. 12, no. 2, Jun. 2018.

[12] P. Boldi, A. Pietri, S. Vigna, and S. Zacchiroli, "Ultra-large-scale repository analysis via graph compression," in *SANER*. IEEE, 2020, pp. 184–194.

[13] A. Pietri, D. Spinellis, and S. Zacchiroli, "The software heritage graph dataset: Public software development under one roof," in *2019 IEEE/ACM 16th International Conference on Mining Software Repositories (MSR)*, 2019.

[14] M. Koohi Esfahani, P. Boldi, H. Vandierendonck, P. Kilpatrick, and S. Vigna, "MS-BioGraphs: Sequence similarity graph datasets," 2023. [Online]. Available: https://doi.org/10.48550/arXiv.2308.16744

[15] T. Fontana, S. Vigna, and S. Zacchiroli, "Webgraph: The next generation (is in rust)," in *ACM Web Conference 2024*, 2024.

[16] T. Feder and R. Motwani, "Clique partitions, graph compression and speeding-up algorithms," in *Proceedings of the twenty-third annual ACM symposium on Theory of computing*, 1991, pp. 123–133.

[17] R. A. Rossi and R. Zhou, "Graphzip: a clique-based sparse graph compression method," *Journal of Big Data*, vol. 5, no. 1, p. 10, 2018.

[18] Q. Abbas, M. Koohi Esfahani, I. Overton, and H. Vandierendonck, "Qclique: Optimizing performance and acuracy in maximum weighted clique," in *Euro-Par 2024*.

[19] S. Navlakha, R. Rastogi, and N. Shrivastava, "Graph summarization with bounded error," in *Proceedings of the 2008 ACM SIGMOD international conference on Management of data*, 2008, pp. 419–432.

[20] F. Zhou, "Graph compression," *Department of Computer Science and Helsinki Institute for Information Technology HIIT*, pp. 1–12, 2015.

[21] Z. Chen, F. Zhang, J. Guan, J. Zhai, X. Shen, H. Zhang, W. Shu, and X. Du, "Compressgraph: Efficient parallel graph analytics with rule-based compression," *Proceedings of the ACM on Management of Data*, vol. 1, no. 1, pp. 1–31, 2023.

[22] B. Ribeiro and D. Towsley, "Estimating and sampling graphs with multidimensional random walks," in *Proceedings of the 10th ACM SIGCOMM conference on Internet measurement*, 2010, pp. 390–403.

[23] W. Fan, J. Li, X. Wang, and Y. Wu, "Query preserving graph compression," in *Proceedings of the 2012 ACM SIGMOD international conference on management of data*, 2012, pp. 157–168.

[24] A. C. Gilbert and K. Levchenko, "Compressing network graphs," in *Proceedings of the LinkKDD workshop at the 10th ACM Conference on KDD*, vol. 124, 2004.

[25] S. Maneth and F. Peternek, "A survey on methods and systems for graph compression," *arXiv preprint arXiv:1504.00616*, 2015.

[26] Š. Čebirić, F. Goasdoué, H. Kondylakis, D. Kotzinos, I. Manolescu, G. Troullinou, and M. Zneika, "Summarizing semantic graphs: a survey," *The VLDB journal*, vol. 28, pp. 295–327, 2019.

[27] A. Khan, S. S. Bhowmick, and F. Bonchi, "Summarizing static and dynamic big graphs," 2017.

[28] T. Liu and D. Li, "Endgraph: An efficient distributed graph preprocessing system," in *2022 IEEE 42nd International Conference on Distributed Computing Systems (ICDCS)*, 2022, pp. 111–121.

[29] F. Sheng, Q. Cao, H. Cai, J. Yao, and C. Xie, "Grapu: Accelerate streaming graph analysis through preprocessing buffered updates," in *Proceedings of the ACM Symposium on Cloud Computing*. ACM, 2018.

[30] M. Then, M. Kaufmann, A. Kemper, and T. Neumann, "Evaluation of parallel graph loading techniques," ser. GRADES '16. ACM, 2016.

[31] Z. Lv, M. Yan, X. Liu, M. Dong, X. Ye, D. Fan, and N. Sun, "A survey of graph pre-processing methods: From algorithmic to hardware perspectives," 2023.

[32] P. Braam, "The lustre storage architecture," 2019. [Online]. Available: https://arxiv.org/abs/1903.01955

[33] A. George, A. Dilger, M. J. Brim, R. Mohr, A. Shehata, J. Y. Choi, A. M. Karimi, J. Hanley, J. Simmons, D. Manno, V. M. Vergara, S. Oral, and C. Zimmer, "Lustre unveiled: Evolution, design, advancements, and current trends," *ACM Trans. Storage*, 2025.

[34] F. Tosoni, P. Bille, V. Brunacci, A. d. Angelis, P. Ferragina, and G. Manzini, "Toward greener matrix operations by lossless compressed formats," *IEEE Access*, vol. 13, pp. 56756–56779, 2025.

[35] R. Meusel, S. Vigna, O. Lehmberg, and C. Bizer, "The graph structure in the web – analyzed on different aggregation levels," *The Journal of Web Science*, vol. 1, no. 1, pp. 33–47, 2015.

[36] O. Lehmberg, R. Meusel, and C. Bizer, "Graph structure in the web: Aggregated by pay-level domain," in *Proceedings of the 2014 ACM Conference on Web Science*, ser. WebSci '14. ACM, 2014, p. 119–128.

[37] R. Meusel, S. Vigna, O. Lehmberg, and C. Bizer, "Graph structure in the web — revisited: A trick of the heavy tail," ser. WWW '14 Companion. ACM, 2014.

[38] P. Boldi, M. Santini, and S. Vigna, "A large time-aware graph," *SIGIR Forum*, vol. 42, no. 2, pp. 33–38, 2008.

[39] H. Kwak, C. Lee, H. Park, and S. Moon, "What is twitter, a social network or a news media?" in *Proceedings of the 19th International Conference on World Wide Web*, ser. WWW '10. ACM, 2010, p. 591–600.

[40] R. C. Murphy, K. B. Wheeler, B. W. Barrett, and J. A. Ang, "Introducing the graph 500," *Cray Users Group*, 2010.